# Using math in physics:
# 7. Telling the story

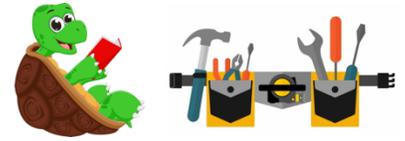

**Edward F. Redish,**
University of Maryland - emeritus, College Park, MD

Even if students can *make the blend*— interpret physics correctly in mathematical symbology and graphs, they still need to be able to apply that knowledge in productive and coherent ways. As instructors, we can show our solutions to complex problems in class. We can give complex problems to students as homework. But our students are likely to still have trouble because they are missing a key element of making sense of how we think about physics: How to tell the story of what's happening.

We use math in physics differently than it's used in math classes.[1] In math classes, students manipulate equations with abstract symbols that usually have no physical meaning. In physics, we blend conceptual physics knowledge with mathematical symbology. This changes the way that we use math and what we can do with it.

We use these blended mental structures to create stories about what's happening (mechanism) and stabilize them with fundamental physical laws (synthesis). Fig. 1 shows the importance of synthesis and mechanism in making sense of the physical world and how all the steps are tied together with equations (with math).

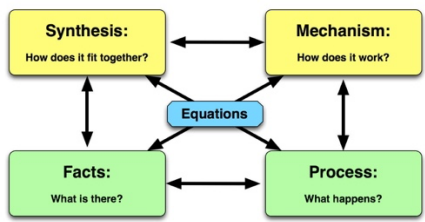

*Fig. 1: Building coherent physics knowledge*

Many students coming into our physics classes need to shift their view of what it means to do science and how to use math in science. Students too often focus only on the bottom boxes of Fig. 1, memorizing facts and process steps (algorithms). We want them to be able to tell a coherent story about the problem. We want them to not just know the answer. We want them to understand what's going on: What's the mechanism? How does this fit with fundamental physics principles (synthesis)?

The research literature uses the term epistemological framing for these different views of what to do to learn — how an individual interprets the nature of the knowledge being learned and what knowledge is needed to deal with a particular situation.[2]

What can we as instructors do to be explicit and help students reframe what it means to use math in science?

In earlier papers in this series I've suggested that a useful way is to identify tools to help students build the blend.[3] The tools I've covered in the previous papers are important and useful. But to solve complex problems, they'll need to learn the essential element: telling the story.

## The key element is how to tell the story

We all know the importance and the stability of stories from our everyday experiences. Well into adulthood, we can easily recount stories we learned as children, whether from Aesop's fables or Disney movies. The moral messages from those stories help build how we function and who we are.

Storytelling is a fundamental resource we all have to organize our long-term memories. We also generate stories to make sense of what's happening in our everyday lives — what's going on, what something means.[4]

Solving a complex physics problem by making sense of it means building a new story, adapted from previously learned stories and drawing on blended concepts and fundamental physical principles.

How can we as instructors help students use their existing storytelling resources to learn to build appropriate stories when solving physics problems?

## Storytelling is a fundamental tool in the mathematical toolbelt

Case studies in the research literature suggest that being explicit about epistemological issues can have a powerful impact, even crossing disciplinary boundaries[5] and lasting beyond the individual physics class.[6]

My approach is to identify telling the story as part of the mathematical toolbelt I've been writing about in this series of papers. As with the other tools, I help students see storytelling as useful by being explicit about its use and expanding the kind of tasks they get in order to focus on learning this skill.

As with the other tools, I assign storytelling an icon: a reading terrapin (the U. of Maryland's mascot) that you see next to the title of this article. I use it on slides, text, and problem solutions whenever the students and I are telling a story to set up a problem, solve it, or decide that we've done it correctly.

In the rest of this paper, I explain what the cognitive science research literature tells us about how memory for stories works — illustrated with my own physics stories. Understanding that research can help you see what our students are going through as they develop and use their storytelling skills for the physics blend. Then, I give five specific techniques for helping students learn to tell and value stories as a way to understand mechanism and synthesis in physics.

## Stories are central to our creation of stable long-term memories

In the early 1990s, when I was just learning about Physics Education Research, I invited Ron Thornton, one of the developers of active-learning supported by computer-assisted measurement devices, to give a seminar to my research group on using those tools. He demonstrated the sonic ranger and claimed that students had serious difficulties reading a velocity graph. He said it was harder than it looks.

To show this, he called for a "volunteer", pointing at me (the senior member of the research group). He showed a velocity graph (Fig. 1A) and asked me to walk to match it with the sonic ranger taking data on my position but displaying my velocity. I had no hesitation in doing this, having taught the concept in introductory physics classes for more than 20 years.

When he started collecting data, I confidently strode backwards, my velocity rising to match the graph. But…when I got the value of my velocity to match, instead of continuing at that velocity as shown by the graph, I stopped! And produced something like the red graph in Fig. 2B.

"Wait! Wait" I said. "I have to put my brain in velocity mode!" The second time I was able to walk the correct graph with no difficulty, making Ron's point perfectly, to the amusement of my grad students.

But the point of this story is not the story itself, but rather a "meta-story" — a story about the story.

Soon after this, I began to put this story into the many seminars I gave introducing PER to physics departments around the country. Some years later, I gave an invited talk at an AAPT meeting.[7] Someone came up to me after the talk, thanked me for it, and commented that they had heard me lecture about PER many years ago — and they most enjoyed the story about the sonic ranger. They were able to recount it correctly more than a decade later!

But…

## The stories we recall aren't always accurate

Another story, this time from my personal experience teaching introductory physics to engineers, illustrates this.

I was one of a number of instructors teaching large lecture classes in introductory physics for engineers in the late 1990s. Between the 1st and 2nd semesters, some number of students would switch from one instructor to another, typically to get lecture times that fit with the rest of their schedule.

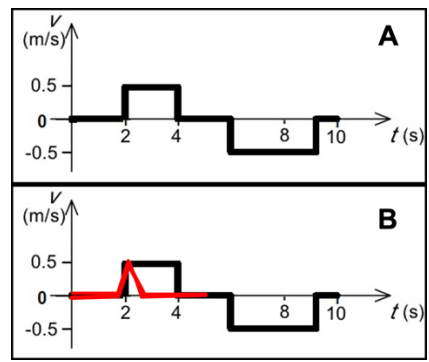

*Fig. 2: Walking a velocity graph – wrong!*

One day, I had done a demonstration in class with a surprising result: a demonstration of Lenz's law (sort of [8]) producing a ring jumping 10 feet in the air. After class, Calvin came up to me and said that he liked my demonstrations much more than the ones his 1st semester prof had done. When I asked, "Why?" he responded, "Well, he just did demos that showed us that what we knew was going to happen, did happen." I asked for an example. Calvin replied, "Like with the racing billiard balls on the straight and dipped tracks." (See Fig. 3.) "We knew that since the ball on the dipped track sped up but slowed back down, they would get to end at the same time, and that's what happened."

Well! The story he told himself was good enough to help him remember the result into the next semester. The only problem is that's not what happens! The ball on the dipped track speeds up as it falls and returns to its original speed as it rises back to the straight part of the track, so the horizontal component of that ball's speed is always equal to or greater than that of the ball on the un-dipped track. As a result, the one on the dipped track reaches the end first. This had to be what Calvin saw. But it was not what he remembered.

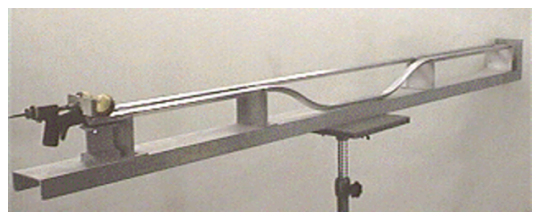

*Fig. 3: Two spring-launched billiard balls race along a straight and a dipped track*

## Stories are reconstructed rather than recorded

A lot of anthropological[9] and cognitive[10] research supports the idea that memories are not just replays of what happens but are put together from knowledge bits and story lines. This actually makes good sense.

The point of memory is not for me to be able to sit on the porch and spin amusing tales of my past to my grandchildren. It's so we can find food, avoid predators, and win a mate to pass on our genes to descendants. These tasks require predicting the future by creating stories about what will happen from what you know about what did happen.

We create these future stories by putting together elements in our memory and making a plausible story about what will happen from what we've experienced. So when we tell a story about a past event, we're not bringing up a stored accurate video. We're "predicting the past".[11,12]

In helping our students learn to tell stories about physics problems, we have to figure out how to help them not just come up with a story, which will often tie to an inappropriate but easy to recall fact or generalization, but to create a correct story that ties to fundamental physics principles. We have to help them to not just tell a hand-waving story, but to tie it to the symbolic math, blending the math with physics meaning.

But first, they have to decide to try to create a story.

## Students have to choose to tell and value a story

At the introductory level, students, especially those who have succeeded in earlier science classes by memorizing answers, often get things wrong by quick associations. "Oh, I remember this. It's…" I call this "one-step thinking" [13] (for students, or "p-primming"[7] for researchers).

If a calculation is required, students may simply reach for an equation and put in numbers inappropriately ("recursive plug-and-chug"[14]), even putting in inappropriate quantities, like a velocity for a volume, if they haven't learned to blend concepts and symbols. Students' belief that they need to (only) get answers in a science class often blocks their building an appropriate story. (Why do they do this? Read the two stories at the beginning of the Supplementary Materials!)

Even if they build a story, they might not know to value it.

My research group observed a dramatic event as we watched (and videotaped[15]) two physics majors, Merry and Pippin, working together on a homework problem in an E&M class from the Griffith's textbook.[16]

A square loop (Fig. 4, top) in the x-y plane is carrying a constant current. There is a constant B-field pointing in the x-direction that varies with y as $\vec{B} = (B_0/L)y\hat{\imath}$. (Fig. 4, bottom) Is there a net magnetic force on the loop? If so, in what direction does it point?[17]

They each finished a solution. Merry had a story. "This looks pretty trivial — like a Physics 2 problem. You just use the right-hand rule. The forces on the sides cancel but the forces on the top and bottom both point down so the answer is down." But Pippin had a calculation. "We're learning about line integrals of vectors, so I think they want us to do the line integral. I worked it out and got 0."

I would have expected Merry to respond something like this: "Well, I can't see how the simple physics could be wrong. You must have dropped a sign somewhere. Let's go through your calculation and find it." Not. What. Happened. Merry immediately folded their hand and copied Pippin's solution.

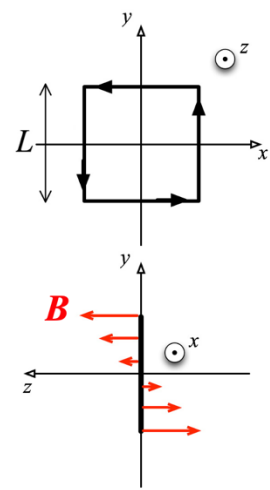

Fig. 4: In what direction is the magnetic force on a current-carrying loop of wire?

This shows both the value of having a good story tied to fundamental physics and the need to value that story. Students need to have and use the epistemological resource that a simple story based on a solid physical principle is a good way to check a complex calculation.[18] Vector integrals are tricky and it's easy to make a sign error!

Putting together Calvin's story with the dipped track with Merry and Pippin's story with the current-carrying loop, we see that success requires not just valuing stories.[19] They have to be tied to fundamental principles, done with a careful consideration of context and the applicability of the principle, and connected to calculations.

What can we as instructors do to help students tell accurate stories, value them, and use storytelling to appreciate and understand physical mechanism and synthesis?

## You can help your students learn to tell useful stories in class

We want

- students to tell a story about what's happening and not just reach for an equation or a remembered answer
- the story they tell to be physically correct (that is, appropriately based on fundamental physical principles)
- that story to be "mathematizable" — tied to blended symbology so as to be able to guide setting up a calculation, moving to a solution, and evaluating its correctness

As professional physicists, we learn that the most important first step in approaching a complex problem is to ask, "What's the physics?" By this, we mean, "What story can we tell that will help us set up an appropriate reasoning chain or calculation, based on fundamental physics principles?"

We also want to help students learn to do this. However, because of all the many ways remembered stories can fail, we don't want to just help students "learn to tell the story of a physics problem". We want them to "learn to tell it in a physically reliable way."

In this section, I give you five techniques that I have found help students build their storytelling tool. It's particularly useful to rely on variation — changing situations and perspectives to see how the results change. This helps students focus on process and what they're doing rather than just the answers.[20]

- Scaffold building a story tied to blended symbols and based on principles.
- Foster coherence by telling a story in graphs with multiple variables.
- Remove the scaffold by presenting problems that require students to both create a story and use fundamental principles.
- Use essay questions to get students to think about their thinking.
- Blend two independent stories — reconcile personal experience (often with one-step thinking) — and physical reasoning (based on fundamental principles).

Learning to tell the story also is a part of learning to use the other tools in the mathematical toolbelt. Examples in previous papers of this series rely on constructing a coherent story.

- In Anchor Equations: Making up a lap[3(d)]
- In Modeling: The pellet, cart, and spring[3(e)]
- In Physics in a Graph: Cart on a tilted air track with spring[3(g)]

## Scaffold building a story tied to blended symbols and based on principles

One way that I particularly like for helping students learn to build the physical story is with problem triples: a qualitative problem, a symbolic problem, and a computational problem. The qualitative problem is tied to symbolic quantities to help them build the blend and to call on fundamental principles. The symbolic problem shows how to use equations in qualitative reasoning and how one can build new equations from the combination. Once the mathematized story is built, setting up the computation is straightforward and makes sense.

Here's an example with a simple toy model: pushing a block resting on a block.

To answer these questions, you have to apply Newton's $1^{st}$ and $2^{nd}$ laws qualitatively to the conceptually blended variables and tell the story of what is happening to each object in the problem. Fig. 5 shows the parts of the qualitative question. (See the Supplementary Materials for the answers to all the problems in the rest of the paper.)

In the second (symbolic) version of the problem, students are asked to find relations among the various forces in the problem and to decide which forces are easily measured (force of the finger, weight of the boxes), and which are "invisible forces" that have to be inferred in each case from applying Newton's laws (normal and frictional forces).

In the third (computational) version, students are asked to calculate various forces given numbers for the masses and coefficients of friction.

One part of the third version is typically the only one students are given in traditional classes, assuming that they can build the necessary story and the resulting relations without guidance, and missing the enlightening comparison of the different situations. (The $2^{nd}$ and $3^{rd}$ versions of the problem are in the Supplementary Materials, with detailed answers displaying how the story is used to support the calculation.)

You can also ask students to explicitly tell the story of what's happening to related variables in a problem, as in the *Model Rocket* problem in the Supplementary Materials.

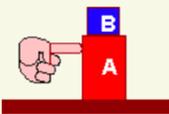

*Fig. 5: A qualitative problem helping students build the story of what's happening in a situation that looks simple, but isn't.*

## Foster coherence by telling a story in graphs with multiple variables

In building physically reliable stories tied to blended variables, it can help to tell the story through different lenses. One way to do this is to generate graphs for multiple variables for a single situation (as described in reference 3(g)). I show an example in Fig. 6.

Many of the examples in ref. 3(g) are of this type.

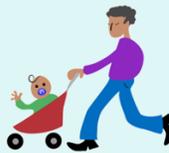

*Fig. 6: Telling a story with multiple graphs*

## Remove the scaffold by presenting problems that require both a story and the use of principles

Fig. 7 shows a multiple choice problem that requires blending, a good development of the story line, and application of fundamental physics principles without providing much scaffolding. (There is some scaffolding in the choice and ordering of the questions.)

To get the full value of this problem, there needs to be a class discussion drawing out how the students solved the problem and what they used to decide on the answers.

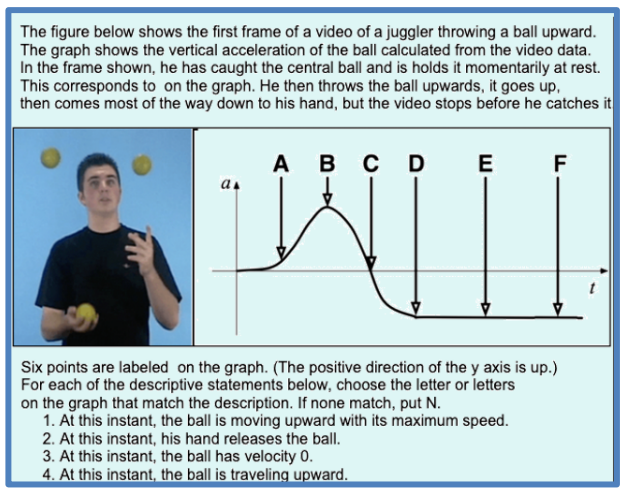

*Fig. 7: A multiple-choice question that requires building a story based on fundamental principles*

## Use essay questions to get students to think about their thinking

It's important to not just use stories tied to fundamental physics principles but to call out explicitly that you are doing so. In class discussions, in asking students to explain their reasoning, and in lecture, explain how you are using stories both to approach creating a solution and to evaluate your solution.

It's also valuable to ask your students to think about and write about the use of stories explicitly. Fig. 8 shows an exam question that I posed for my class one year.[21]

> Emma and Ryan were studying for their Physics exam. Ryan says, "Prof. Redish says that one of the tools we're supposed to have available for solving problems is *telling the story*. How can I 'tell a story' of a physics problem?" Emma had a good explanation and an example. What do you think she might have said? *Note: This is an essay question. Your answer will be judged not solely on its correctness, but for its depth, coherence, and clarity.*

*Fig. 8: An essay question to get students to think about the value of telling the story*

More than half of the students provided satisfactory answers, even under the pressure of an in-person 50-minute hand-written exam. Here's one answer that I was particularly happy with.[22]

> *Analogous to a story, a physics problem has various components. Depending on the context, it'll have different relevant characters who will interact through the story's timeline. For example, if a physics problem involves a model rocket, a fuse, and the earth, we can think of them as our story's characters, each behaving & interacting with one another in a particular way… We could see the fuse behaving so that it influences the rocket, changes its behavior, and resisting the influence of the earth. This could be considered conflict, which climaxes and resolves as the rocket rises, falls, and lands back down.*

I provide another essay question, *Is the momentum of the train conserved?*, asking students to think about how the story of a particular problem is tied to fundamental principle at the end of the Supplementary Materials.

## Blend two independent stories — reconcile personal experience and physical reasoning

The erroneous stories that students spontaneously generate in response to a physics question are often built from quick associations with everyday experience. While these are often referred to as "misconceptions", their stability often comes from the ease of bringing them up,[13] not necessarily from coherent incorrect theories or mental models. If the question is phrased differently, a different story can be activated and a more correct one built and stabilized.

A useful tool for doing this is the paired-question technique developed by Elby.[23] In this approach, instead of offering students introductory questions that research has shown most will get wrong, a pair of matched questions are created so that most students will get one right and one wrong, producing a conflict.[24] An example, shown in Fig. 9, helps reconcile the automatically-generated story that a smaller object hitting a larger must feel a larger force (the p-prim "more means more") with Newton's $3^{rd}$ law that says the two forces must be equal.[25]

> a) A truck rams into a parked car. Intuitively, which is larger: the force exerted by the truck on the car or by the car on the truck?
>
> (b) Suppose the truck has mass 1000 kg and the car has mass 500 kg. During the collision, the truck slows by 5 m/s. How much speed does the car gain?

*Fig. 9: An "Elby pair" question to help students reconcile spontaneous with physics-accurate stories*

Elaboration and analysis of the pair of answers shows students that their intuitions are leading them into a conflict. They are then guided to find their (essentially correct) "raw intuition" that underlies both their answers and are guided to maintain that intuition and to refine it in a way that leads to results that are consistent with the physics they are learning.

In this way, we hope to convince students that their intuitions about the physical world are valuable and, when properly refined, support the physics knowledge they are learning by tying the story to a fundamental principle. Fig. 10 shows an example — an unbalanced force goes with acceleration, not velocity, by the fundamental principle of Newton's 2nd law.

You can see more examples of all these types of questions in the Open Source Tutorials available on *PhysPort*.[26]

## Digging deeper

Many researchers have studied the issues I've explored in this paper. For broader, more complete research-based frameworks, see Kaldara & Wieman[27] or Odden & Russ.[28] Others provide tools for studying student success in telling stories (though they use different terms such as sense-making with mathematical equations,[29,30] following and creating reasoning chains,[31] or developing mechanistic reasoning.[32,33]

A great example of using storytelling is Energy Theater, developed for helping pre- and in-service teachers build the blend for the complex concept of energy.[34] And since modeling usually starts with "finding the physics," the modeling group has links to a lot of resources for teaching modeling that require story building.[35]

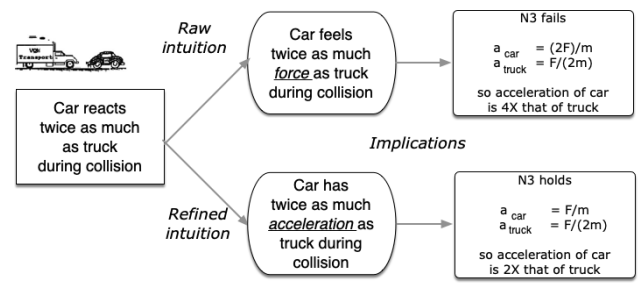

*Fig. 10: Reconciling the conflicting stories generated by Elby pair questions*

## Acknowledgements

I would like to thank the members of the UMd PERG over the last two decades for discussion on these issues. The work has been supported in part by a grant from the Howard Hughes Medical Institute and NSF grants 1504366 and 1624478.